# Vectorial wavefront holography based on a polarisation-insensitive hologram

Haoran Ren[1]

[1] MQ Photonics Research Centre, Department of Physics and Astronomy, Macquarie University, Sydney, Australia.

E-mail: haoran.ren@mq.edu.au

**Abstract**

Polarisation holography generally demands polarisation-sensitive holograms for reconstructing either polarisation-multiplexed holographic images or polarisation-sensitive image channels. To date, polarisation holography is underpinned by the Jones matrix method that uses birefringent holograms, including ultrathin metasurface holograms, limiting the polarisation control to orthogonal polarisation states. Here I introduce a novel concept of vectorial wavefront holography by exploiting the wavefront shaping of a structured vector beam. I will show that a phase hologram can be used to tailor the polarisation interference of a vector beam in momentum space, creating arbitrary polarisation states that include but not limited to the linear, circular, azimuthal, and radial polarisations. This opens an unprecedented opportunity for the multiplexing generation of arbitrary polarisation distributions in a holographic image. The demonstrated vectorial wavefront holography offers flexible polarisation control without using birefringent optical materials, which may find applications in polarisation imaging, holographic encryption, holographic data storage, multi-view displays, holographic Stokesmeter, and polarimetry.

## 1. Introduction

Optical holography provides a disruptive technology that allows the use of a hologram to display 3D optical fields via the complex-amplitude modulation of an incident beam. In addition to the amplitude and phase degrees of freedom, light can also carry polarisation information—the oscillating direction of an electromagnetic light field. Conventional polarisation holograms have been optically recorded in birefringent media that have a polarisation-dependent sensitivity. Illuminating the birefringent media with two interfered orthogonal polarised fields allows the reocrding and reconstruction of polarisation-sensitive holographic images [1,2]. As such, birefringent holograms open the door to polarisation holography that has the potential to increase the bandwidth and security of an optical hologram. However, a long exposure of polarised light into polarisation-sensitive media is required to induce sufficient birefringence, accompanied with an undesired high-temperature treatment. These limitations have hindered the conventional polarisation holography for practical applications.

Recent development of metasurface technology has transformed the field of digital holography, opening the possibility to digitise polarisation-sensitive holograms using birefringent meta-atoms [3–13]. Birefringent metasurface holograms have recently been developed to either carry independent image channels for an improved hologram bandwidth [3,5–7,9–11,14], or reconstruct and display a polarisation-multiplexed distribution [4,8,13]. However, birefrigent metasurfaces underpinned from the Jones matrix method have obvious limitations: for instance, the polarisation access is typically limited to orthogonal polarisations [3,5–7,9]; detour [4,13] or supercell meta-atoms [6,10,11] designed for the polarisation sensitivity sacrifice hologram resolution.

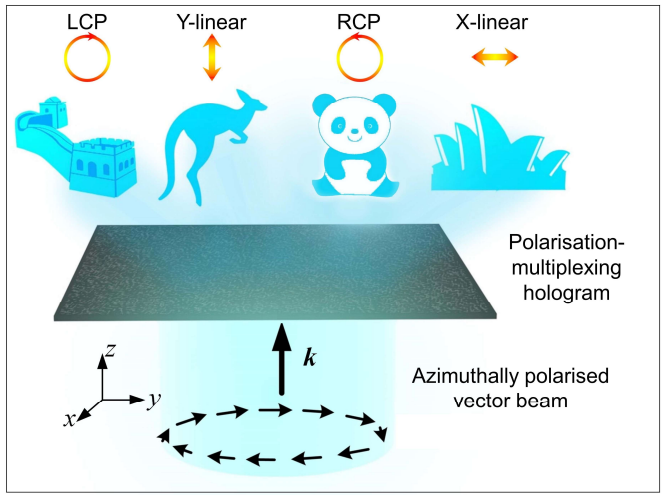

**Figure 1. Principle of vectorial wavefront holography based on the phase-only modulation of an azimuthally polarised vector beam.** The phase-only polarisation-multiplexing hologram offers the simulatenous reconstruction of different polarisation distributions in a holographic image, including both the linear and circular polarisations.

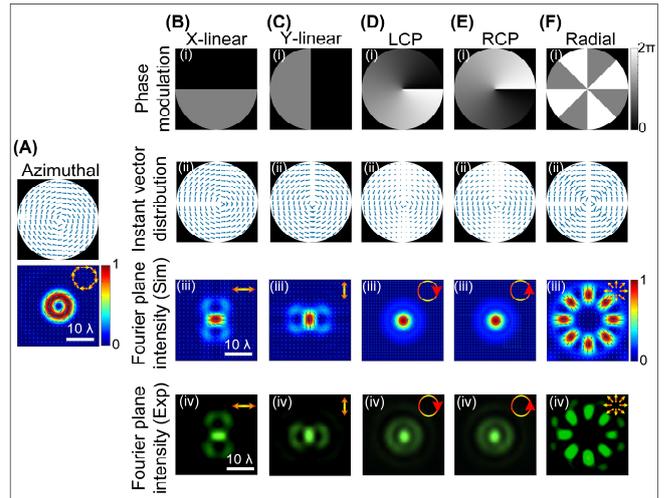

**Figure 2. Generation of arbitrary polarisations based on the phase modulation of an APVB.** (**A**) Instant vector distribution and Fourier plane intensity distribution of an APVB. (**B-F**) The shase maps (i), instant vector distributions (ii), numercial (iii) and experimental Fourier plane intensity distributions (iv) of created different polarisations, including the x-linear (B), y-linear (C), LCP (D), RCP (E) and radial polarisations (F), respectively. The results were based on an incident wavelength of 532 nm, a beam diameter of 8 mm and a Fourier lens with a focal distance of 200 mm.

Here I demonstrate a new concept of vectorial wavefront holography through the phase manipulation of a structured vector beam. Breaking the rotational symmetry of an azimuthally polarised beam via a phase-only hologram allows flexible tailoring of the interference of vector fields in momentum space (the Fourier domain of a hologram), offering a new way for polarisation control. As such, vectorial wavefront holography alleviates the necessity of polarisation-sensitive materials for polarisation holography, making the polarisation control more flexible and robust. I will present the simultaneous reconstruction of different polarisation distributions that include the linear, circular, radial, and azimuthal polarisations in a holographic image. This was achieved from illuminating an azimuthally polarised vector beam (APVB) on a phase-only polarisation-multiplexing hologram (Fig. 1).

## 2. Design principle

The principle of creating arbitrary polarisation states based on the phase-only modulation of a structured vector beam is illustrated in Fig. 2. Unlike conventional optical beams with a homogeneous polarisation distribution, such as a Guassian beam, an APVB, which belongs to a cylindrical beam, has spatially variant electric field vectors perpendicular to the cylindrical beam's radial axis (Fig. 2A). Owing to its polarisation singularity, the APVB features a donought intensity distribution in the momentum space (e.g., on a Fourier plane). Since azimuthal polarisation is immune to the depolarisation effect, no matter how tight the Fourier lens focus is [15,16], it represents an advantagous over other vector beams for pure polarisation manipulation. Nevertheless, other structured vector modes (e.g. radial polarisation) can also be used for the vectorial wavefront holography.

The capability of producing different polarisations through the use of a single phase-modulated APVB was explored in Figs. 2B-2F. To numerically simulate the interference of vector fields in momentum space, vectorial Debye diffraction theory was employed [16,17]. In the simulation, an APVB with a transerve cross-section size of 8 mm, a low numerical aperture Fourier lens with a focal length of 200 mm, and a wavelength of 532 nm were considered. Without loss of generality, five different phase maps were imparted on an APVB for the generation of x-linear (Fig. 2B), y-linear (Fig. 2C), left-handed circular (Fig. 2D), right-handed circular (Fig. 2E), and radial polarisations (Fig. 2F). Specifically, to create linear polarisation, a π-phase-step map was used to break the rotational symmetry of the APVB in the transverse cross-section plane. This allows the electric field vectors aligned with the phase-step line to be constructively interfere in the momentum space, leading to the generation of an arbitrary linear polarisation with a different polarisation axis. The polarisation axis can be controlled by the π-phase-step line (Figs. 2B and 2C). In addition, vortex phase maps with topological charges of +1 and -1 can be imparted on an APVB to create left- and right-handed circular polarisations (LCP and RCP), respectively (Figs. 2D and 2E). The last example is radial polarisation, an another commonly used cylindrical polarisation. Implementing a multi-zone π-phase map with multi-fold (e.g., four-fold) rotational symmetry onto an APVB, a radially polarised vector beam can be produced (Fig. 2F), wherein the number of radially polarised intensity nodes is determined by the rotational symmetry of the multi-zone π-phase map. Consequently, I experimentally verified the generation of arbitrary polarisations through using a spatial light modulator (SLM) to implement the phase maps on an APVB, showing quite consistent results with the simulation ones, as presented in the last row in Figs. 2B-2F.

The use of phase maps for the control of different polarisations provides an alternative approach to polarisation holography, capable of multiplexing generation of different polarisation distributions in a holographic image. The design of a polarisation-multiplexing hologram is illustrated in Fig. 3. According to Fourier optics, when a phase-modulated APVB is impinged onto a Fourier hologram, physical properties of incident light, including amplitude, phase, and polarisation, will be mathematically regarded as the impluse function of Fourier transform (FT) and convoluted to each spatial-frequency component of the hologram. Owing to the FT relation, spatial-frequency components of a Fourier hologram are simply represented by the pixels of the holographic image in momentum space [18–20]. To preserve the polarisation property in momentum space, spatial-frequency components of the hologram need to be sampled, equivalent to the sampling of holographic images by using a two-dimensional (2D) Dirac comb function (Fig. 3). As long as the sampling period is larger than the diffraction-limited impluse response of the phase-modulated APVB in momentum space, the polarisation property of the incident beam can be well preserved in each pixel of the holographic image. This lays the physical foundation of a new concept of vectorial wavefront holography.

Here I demonstrate the parallel generation of different polarisation distributions in a holographic image. To this purpose, a polarisation-multiplexing hologram was designed through the linear superposition of four different Fourier holograms (Fig. 3). Each polarisation image channel was firstly sampled by a 2D comb function, the result of which was used as the image target for the phase retrieval of a Fourier hologram through an iterative optimisation algorithm [15,16]. To produce different polarisation distributions in a holographic image, each Fourier hologram was further encoded with a distinctive phase map for the polarisation control. For a better visualisation, a grating phase map was further added to each Fourier hologram for spatially separating the encoded holographic image channels.

Mathematically, a polarisation-multiplexing hologram can be represented as superposition of complex-amplitude fields of different polarisation image channels encoded with distinctive polarisation-controlling phase maps in the hologram plane: $E^{mul} = \sum_{j=1}^{M} e^{i\phi_j} e^{i\varphi_j}$, wherein $\phi_j$ and $\varphi_j$ stand for the phase-retrived hologram and the distictive polarisation-controlling phase map for each polarisation image channel, respectively; $M$ denotes the total number of multiplexing channels. Since the complex-amplitude hologram is Fourier-based, its reconstructed optical fields in momentum space can be represented as $\sum_{j=1}^{M} \mathcal{F}(e^{i\phi_j}) \otimes \mathcal{F}(e^{i\varphi_j})$, where $\mathcal{F}$ denotes the FT operator, expressing multiplexing results as the superposition of the convolution of each holographic image channel and its phase-controlled polarisation property. For a phase-only hologram, the polarisation-multiplexing hologram can be further described as an argument result: $P^{mul} = \arg[\sum_{j=1}^{M} e^{i\phi_j} e^{i\varphi_j}]$.

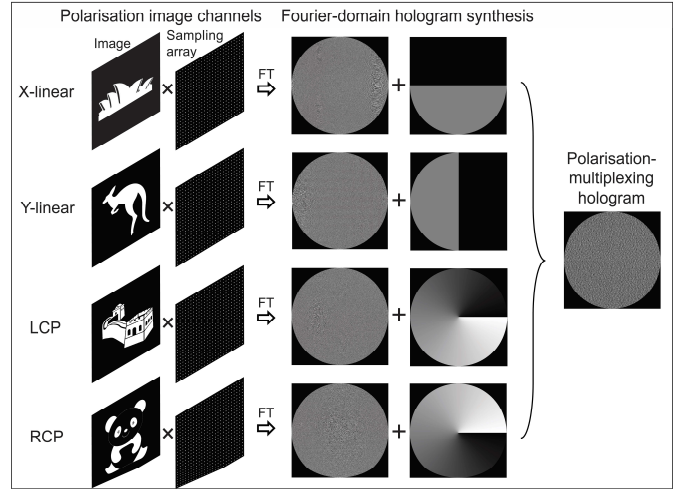

**Figure 3. The flowchart of designing a polarisation-multiplexing hologram.** Four image channels of "Sydney Opera House", "Kangaroo", "The Great Wall", and "Panda" were encoded with the x-linear, y-linear, LCP and RCP polarisations, respectively. To preserve the polarisation information, original image targets were sampled by a 2D comb function. The sampled images were used as new targets for the phase retrieval of Fourier holograms via a Fourier transform (FT). Fourier holograms were further added distinctive polarisation-controlling phase maps for creating different polarisation distributions. Superposition of all the complex-amplitude Fourier holograms leads to the design of a polarisation-multiplexing hologram.

## 3. Experimental results

To experimentally verify the vectorial wavefront holography concept, the designed polarisation-multiplexing hologram was experimentally implemented through a SLM. An azimuthal polarisation converter (APC) was used to convert an incident linear polarisation into an APVB. The schematic diagram of the optical setup is shown in Fig. 4A. A Fourier lens with a focal distance of 200 mm was used to perform the Fourier holography, giving rise to a polarisation-multiplexed holographic image in momentum space, which was detected by a charge-coupled device (CCD). To verfiy different polarisation distributions, a linear polariser (LP) or the cascaded use of a quarter wave plate (QWP) and a LP have been utilised to distinguish either the linear or circular polarisations, respectively. As a result, the parallel reconstruction of four different polarisation distributions in a holographic image was experimentally verified (Fig. 4B). By passing the holographic image through a LP, polarisation image channels of "Sydney Opera House" and "kangaroo" were identified to possess x-linear (Fig. 4C) and y-linear (Fig. 4D) polarisations, respectively. In addition, passing the holographic image through the combination of a QWP and a LP, polarisation image channels of "The Great Wall" and "panda" were detected to carry LCP (Fig. 4E) and RCP (Fig. 4F), respectively. Therefore, vectorial wavefront holography allows the reconstruction of an arbitrary polarisation distribution, which is not limited to orthgonal polarisations, paving the way for advanced and flexible polarisation hologrpahy without using birefrigent optical materials.

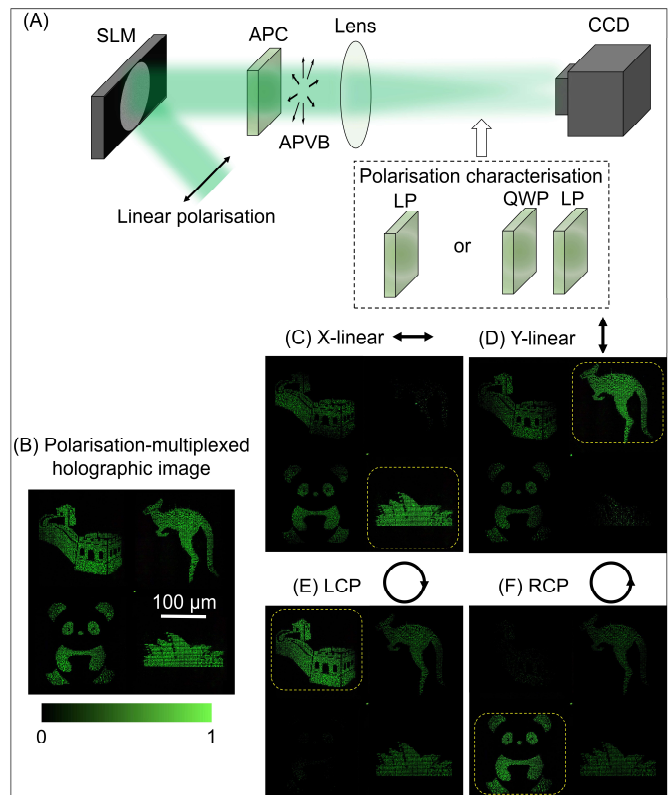

**Figure 4. Experimental demonstration of vectorial wavefront holography.** (**A**) A schematic diagram of the optical setup used for both implementation of vectorial wavefront holography and verification of different polarisation distributions in a reconstructed holographic image. SLM: spatial light modulator, APC: azimuthal polarisation converter, LP: linear polariser, QWP: quarter wave plate. (**B**) Experimental reconstruction of a polarisation-multiplexed holographic image captured by a CCD. (**C**-**F**) Polarisation filtered holographic images for distinguishing the x-linear (C), y-linear (D), LCP (E) and RCP (F) polarisations, respectively. The yellow dotted lines label out specific image channels with target polarisations.

## 4. Conclusions

In conclusion, I have introduced a new concept of vectorial wavefront holography based on the phase-only modulation of a structured vector beam. Parallel reconstruction of an arbitrary polarisation distrbution (including non-orthogonal polarisations) in a holographic image has been numerically designed and experimentally verified. Instead of developing polarisation-sensitive birefrigent optical materials for polarisation holography with a limited polarisation access, this paper applies simple phase maps for tailored polarisation interference of a structured vector beam in momentum space. This alleviates the necessity of polarisation-sensitive materials for polarisation holography. Despite the fact that azimuthal polarisation exhibits a high degree of polarisation purity for the momentum-space polarisation manipulation, vectorial wavefront holography can be realised from other structured vector beams. Indeed, the rich family of structured vector beams offers a trememdous resource and great flexibility for vectorial wavefront holography [21]. Since the orbital angular momentum of light can be solely controlled by a helical phase [22–24], this work can be further extended to manipulate both the polarisation and orbital angular momentum distributions in a holographic image. Therefore, I believe this demonstration will advance holography technology and may find signficant impacts on a multituide of photonic applications, such as polarisation imaging [25], holographic encryption [20], polarisation-encoded data storage [15], multi-view displays [26], holographic Stokesmeter [27], and polarimetry [28].

## Acknowledgements

H. R. acknowledges the funding support from the Macquarie University Research Fellowship (MQRF) from Macquarie University.